\documentclass[useAMS,usenatbib]{mn2e}
\usepackage{graphicx}
\usepackage{amsmath,amssymb}
\newcommand{\del}{\partial}

\newcommand{\f}{\frac}

\newcommand{\BF}{\begin{figure}\begin{center}}
\newcommand{\EF}{\end{center}\end{figure}}
\newcommand{\BE}{\begin{equation}}
\newcommand{\EE}{\end{equation}}
\newcommand{\BEA}{\begin{eqnarray}}
\newcommand{\EEA}{\end{eqnarray}}

\begin{document}
\title{On the origin of the Cold Spot }
\author[Kaiki Taro Inoue]{Kaiki Taro Inoue$^{1}$\thanks{E-mail:
kinoue@phys.kindai.ac.jp}\\
$^{1}$Department of Science and Engineering, 
Kinki University, Higashi-Osaka, 577-8502, Japan   }
\date{\today}

\pagerange{\pageref{firstpage}--\pageref{lastpage}} \pubyear{0000}

\maketitle

\label{firstpage}
\begin{abstract}
In a concordant $\Lambda$ Cold Dark Matter ($\Lambda$CDM) model, large-angle 
Cosmic Microwave Background (CMB) 
temperature anisotropy due to linear perturbations in the local universe 
is not negligible. We explore a possible role of an underdense region (void)
that may cause an anomalous Cold Spot (CS) in the CMB sky.
Although the observed anomalous cold region with a surrounding 
hot ring can be produced by an 
underdense region surrounded by a massive wall, a decrement
in the CMB temperature in the line-of-sight 
is suppressed because of blueshift of CMB 
photons that pass the wall. Therefore, 
undercompensated models give better agreement with
the observed data in comparison with 
overcompensated or compensated models.
We find that it is likely that 
$\sim$90 per cent of the CMB fluctuation is generated due to 
an overdense region surrounded by an underdense region at the 
last scattering surface, and the remaining $\sim$10 per cent 
is produced due to a single spherical  
underdense region with a radius $r\sim 6\times
 10^2~ h^{-1}$Mpc and a density contrast 
$\delta_m\sim -0.009$ ($2 \sigma$) 
at redshift $z\sim 1$ in the line-of-sight to the CS. The probability of
having such two aligned structures is $\sim 0.7$ per cent if
the perturbed region at $z\sim 1$ is moderately undercompensated.   
\end{abstract}

\section{Introduction}
One of the interesting non-Gaussian 
features found in the cosmic microwave background
(CMB) sky is the presence of the Cold Spot (CS) in the Southern 
hemisphere \citep{Cruz05}. Although the statistical significance is 
not significantly high (1.85 per cent) \citep{Cruz07a}, it maybe 
a hint of new physics that is absent in 
the concordant $\Lambda$CDM model. The origin of the 
temperature decrement may be a time-varying
gravitational potential of a supervoid with radius $200-300h^{-1}$Mpc 
in the line-of-sight to the CS \citep{Inoue06, Inoue07},
or that of a cosmological texture at high redshifts $z>1$ \citep{Cruz07b}.
So far, no clear evidence of supervoids in the line-of-sight has
been obtained \citep{Rudnick07, Bremer10,Granett10}. 
If such a huge supervoid exists, it will affect various kinds of  
observables \citep{Das09, Masina09a, Masina09b}. 
Therefore, it can be easily distinguished 
from other scenarios such as
a cosmic texture \citep{Cruz07b} in the near future. 

A key issue is an evaluation of the 
effect of non-compensation of mass. 
It has been pointed out that the CS is anomalous 
because of a hot ring-like structure around it \citep{Zhang10, IST10}. 
However, producing a hot ring
by a linear or quasi-linear compensated 
void is difficult due to the linear integrated Sachs-Wolfe (ISW) 
effect though such a feature 
can be realised by non-linear compensated 
clusters or voids \citep{Tomita08,Sakai08,IST10,Hu11}. 
If one considers a massive wall 
that overcompensates the low density region inside, 
such a feature might be easily made even in the case of
linear fluctuations. However, we need to estimate
the chance of having such structures. 

Moreover, in order to explain the entire feature of the 
CS by a single supervoid, one has to assume 
an exceptionally huge void, which is difficult to 
produce in the standard cosmological scenarios \citep{Inoue07}.
In fact, it is more natural to assume that most of the 
feature of the CS is produced by fluctuations in the gravitational
potential at the last
scattering surface \citep{Valkenburg12}. 
Therefore, one must take 
into account various effects such as the ordinary Sachs-Wolfe (OSW) effect (due to spatial 
fluctuations of the gravitational potential and the temperature) 
and the Doppler effect due to the bulk velocity of photon-baryon fluid
at the last scattering epoch.
It is of crucial importance to study the expected
value of the percentage of the contribution
from an assumed supervoid (or a low density region) to the CS
and the most probable redshift in the framework of 
standard scenarios that predict Gaussian primordial
perturbations. Even if the CS has been produced as a fluke,
an estimation of a possible contribution from the local
structures in the line-of-sight to the CS is important 
since the ISW contribution is expected to be maximum in that direction. 
 
The present paper is organised as follows.
The linear and the non-linear 
ISW effects due to an non-compensated void (or cluster)
are explored in section 2.
The statistical significance of 
a moderately undercompensated underdense region that will partially 
explain the CS is discussed
in section 3. We conclude the paper in section 4.
In the following, unless noted, we assume a 
concordant $\Lambda$CDM cosmology with $(\Omega_{m,0}, \Omega_{\Lambda,0}, \Omega_{b,0}, h,\sigma_8,n)=(0.26,0.74,0.044,0.72,0.80,0.96)$, which agrees with the recent CMB
and large-scale structure data \citep{Sanchez09}.

\section{Model of non-compensated void/cluster}
In order to investigate temperature
variation due to non-compensated underdense or overdense regions,
we consider spherically symmetric density
fluctuations with a comoving radius $r_0$ surrounded by a wall
at $r_0 < r \le r_1$. To localise the perturbation, we require that 
the perturbation is globally compensated.  To do so, we introduce
a cutoff comoving radius $r_2 \ge r_1$ where the total mass of the
perturbation is 
compensated if integrated over $0<r<r_2$.
In what follows, for simplicity, 
we only consider top-hat type matter density perturbations for void (cluster) 
whose linear density contrast is given by
\BE
 \delta^L=\left\{
         \begin{array}{l l l}
           \delta_0^L , & r<r_0 \\
           \delta_1^L , & r_0\le r<r_1 \\
           \delta_2^L , & r_2\le r,\\
         \end{array}
          \right.
\EE
where $\delta_0^L,\delta_1^L$ and $\delta_2^L$ are constant (Fig.
1). From a condition that the perturbation is local, we have
$\delta_0^L r_0^3+\delta_1^L (r_1^3-r_0^3)+\delta_2^L(r_2^3-r_1^3)=0$.
To describe the local mass compensation of the perturbation, 
we define a parameter $\varepsilon$ as 
\BE
\varepsilon\equiv \frac{\delta_0^L r_0^3+\delta_1^L (r_1^3-r_0^3)}{
|\delta_0^L r_0^3|}.\EE
$\varepsilon=0$ corresponds to a locally compensated case, whereas
$\varepsilon >0, (\textrm{or}~\varepsilon <0)$ corresponds to a locally 
overcompensated (undercompensated) case in the region defined by $0\le r\le r_1$ in
linear order.  Even in the case $\varepsilon=0$, if  $r_1-r_0 \gtrsim r_0$
and $r_1=r_2$, we call it a '' moderately over(under)compensated'' model in 
the following\footnote{Alternatively, 
we may call it ``a model with a thick wall''.}. 
\begin{figure}
\begin{center}
   \hspace{-0.4cm}
    \includegraphics[width=8cm,clip]{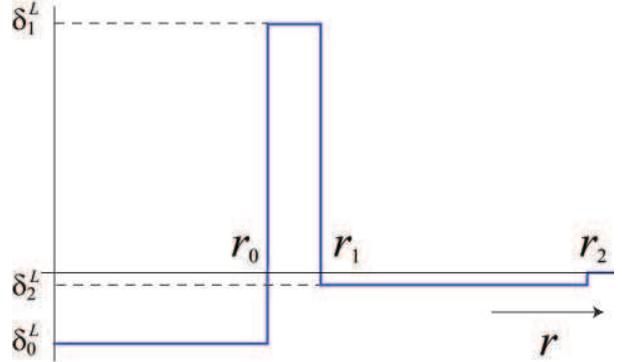}
   \label{figure1}
\end{center}
 \caption{Linear matter density contrast for a top-hat type spherical
 underdense region.}
\end{figure} 

In order to incorporate non-linearity of the scalar-type 
perturbation, we use second order perturbation theory \citep{Tomita08}.
In what follows, we assume that the cosmological Newtonian approximation 
holds for the perturbation that we consider. In other words, we assume
that the
peculiar velocity is sufficiently smaller than the light velocity, and
the size is sufficiently smaller than the present horizon
scale. In what follows we only consider growing mode of density
perturbation that is adiabatic. 
The temperature anisotropy due to time-evolving gravitational potential
perturbation $\psi$ is given by an integration along with a geodesic
of the CMB photon that passes through the potential, 
\BE
\bigg(\frac{\Delta T}{T} \biggr)=2 \int \frac{\del \psi}{\del \eta} d\eta,
\EE 
where $\eta$ is the conformal time. If the amplitude of the 
density contrast $\delta$ is sufficiently small, the 
gravitational potential can be expanded as a sum of linear order and
second order perturbations $\psi=\psi^L+\frac{1}{2}\psi^S$. Time evolution of 
the linear perturbation $\psi^L$ is written as a product of a potential 
function $F$ of comoving coordinates $\bf x$ and a growing factor $D(\eta)$,
 which is a function of the scale factor $a(\eta)$ as
\BEA
\psi^L(\eta,{\bf x})
&=&-D(\eta) F({\bf x})
\nonumber
\\
&=&
\frac{1}{2} \Bigl(1 - \frac{a'}{a}P' \Bigr) F({\bf x}), 
\EEA
where  
\begin{eqnarray}
P(\eta) &=& -{2 \over 3\Omega_{m0}}a^{-3/2} [\Omega_{m0}+\Omega_{\Lambda 0} 
a^3]^{1/2} \nonumber
\\
&\times &
\int^a_0 d a~ a^{3/2}[\Omega_{m0}+\Omega_{\Lambda 0}
 a^3]^{-1/2} +
{2 \over 3\Omega_{m0}}a. 
\end{eqnarray}
Here, $\Omega_{m0},\Omega_{\Lambda 0}$ are density parameters of
non-relativistic matter and a cosmological constant $\Lambda$
of the background FRW universe and a prime denotes derivative with
respect to the conformal time $\eta$. In a similar manner, the second order
perturbation $\psi^S$ can be written as
\BE
\psi^S(\eta, {\bf x}) = \xi_1(\eta)  F_{,i}({\bf x})F_{,i}
({\bf x}) + 100\, \xi_2(\eta) \cdot \Psi_0({\bf x}), 
\EE
where $F_{,i}$ is $\partial F/\partial x^i$, and
\BEA
\xi_1  &=& {1 \over 4}P \Bigl(1 - {a' \over a}P'\Bigr),
\\ 
\xi_2  &=& \Big\{{1 \over 21} {a' \over a}\Bigl(PP' - {1 \over
6}Q'\Bigr) - {1 \over 18}\Bigl[P + {1 \over 2}(P')^2\Bigr] \Big\},
\EEA
and $\Psi_0$ and $Q(\eta)$ satisfy
\BEA
\Delta \Psi_0 &=& {9 \over 200} \Bigl[F_{,ij} F_{,ij} - (\Delta
F)^2\Bigr],
\\
Q'' + {2a'\over a} Q' &=& - \Bigl[P - {5\over 2}(P')^2 \Bigr],
\EEA
respectively (Tomita \& Inoue 2008).
\begin{figure}
\begin{center}
\hspace{-1.1cm}
\includegraphics[width=9.5cm,clip]{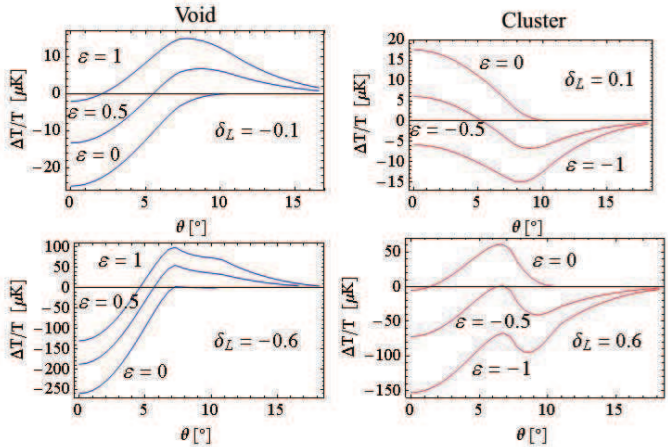}
   \label{figure2}
\end{center}
\vspace{-0.5cm}
 \caption{Effect of noncompensation: temperature anisotropies as a function
 of angular radius $\theta$ from the center of the perturbation at redshift 
$z=1$ are plotted for 
voids (left) and clusters (right). The parameters of
the mass distribution are set to $(r_0,r_1,
 r_2)=(0.1H_0^{-1},0.15H_0^{-1},0.3H_0^{-1})$. Note that the vertical 
scales of the upper and lower panels are different. }
\end{figure} 
First, we consider effects of linear perturbations. 
As shown in Fig.2,  a hot ring that surrounds a cold spot 
appears as the wall around an underdense region becomes 
more massive than a mass deficiency inside
since a decay of gravitational potential 
at an accelerated epoch leads to an increment in the temperature of the
CMB photons.
Similarly, a cold ring that surrounds a hot spot is produced as an
underdense region surrounding an overdense one becomes less massive than
a mass that compensates the overdense region inside. 
These behaviors agree with the 
predicted temperature variation due to the linear ISW effect. 
\begin{figure}
\begin{center}
   \hspace{-0.4cm}
    \includegraphics[width=9cm,clip]{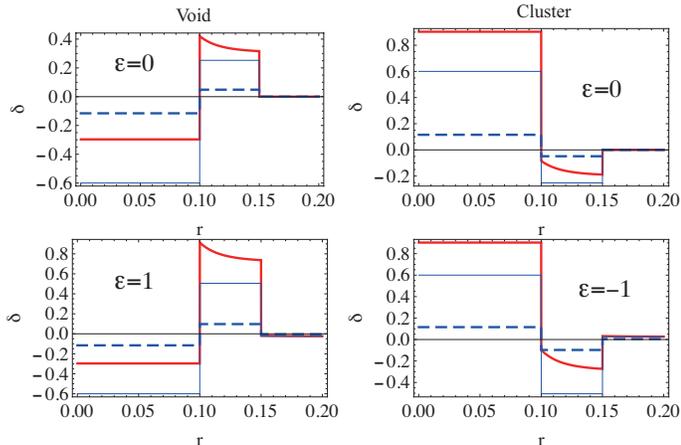}
   \label{figure3}
\end{center}
\vspace{-1cm}
 \caption{Effect of noncompensation:
profiles of 
mass density contrasts for non-linear density perturbations at $z=1$ are
 plotted as thick curves and those for 
corresponding linear density perturbations 
are plotted as thin curves ($z=1$) and
dashed curves ($z=10$) as a function of comoving distance $r$ from the
 center. The unit of the length is the present Hubble radius $H_0^{-1}$.  
The linear density contrasts at the center 
at $z=1$ are assumed to 
be either $\delta_0^L=-0.6$ (left panels) or $\delta_0^L=0.6$
 (right panels). The mass is either locally compensated (upper panels), overcompensated (lower left) or
 undercompensated (lower right). The parameters of
 the mass distribution are set to $(r_0,r_1, r_2)=(0.1H_0^{-1},0.15H_0^{-1},0.3H_0^{-1})$.    
 }
\end{figure} 
\begin{figure}
\begin{center}
   \hspace{-0.2cm}
    \includegraphics[width=8.7cm,clip]{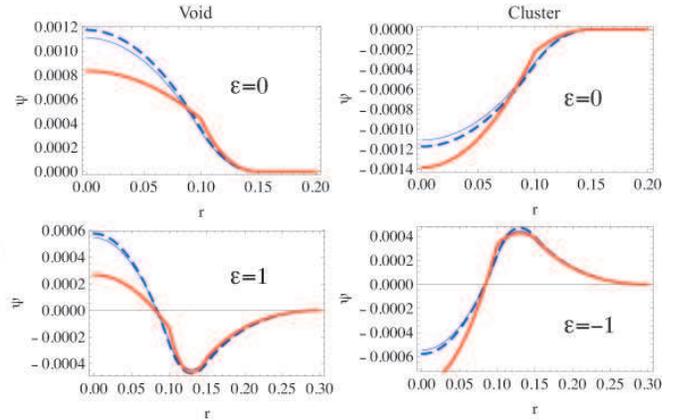}
   \label{figure4}
\end{center}
\vspace{-0.2cm}
 \caption{Effect of noncompensation:
profiles of gravitational potential 
for non-linear density perturbations at $z=1$ are
plotted as thick curves and those for 
corresponding linear density perturbations 
are plotted as thin curves ($z=1$) and
dashed curves ($z=10$) as a function of comoving distance $r$ from the
center. The parameters are the same as in figure 3. }
\end{figure} 

Next, we consider quasi-linear regime. Although the
non-linear mass seems to be overcompensated at $r=r_2$ with
$\varepsilon=0$ (see Fig.3), it is actually locally compensated. 
As shown in Fig.4, the spatial derivative of the 
potential is not continuous at $r_1$. Therefore, it gives a contribution
that is proportional to Dirac's delta function $\delta_D(r-r_1)$.  
Furthermore, if it is smoothed out in the neighbourhood of a point
$r=r_1$, 
the gravitational potential becomes concave outward. For this reason,
the neighbourhood of $r=r_1$ contributes a negative mass to the density
distribution, which leads to a shift of the 
wall towards the outer (inner) shell $r>r_2$ in the case of a void
(cluster).  On the other hand,
an additional mass due to the 
second order effect leads to a decrease in 
the gravitational potential in the inner low density region  
$0 \le r< r_1$. Thus, the gravitational potential expands (shrinks) faster
than the corresponding linear potential in a void (cluster) in the 
outer region, but the value of the potential decreases in the inner
region as shown in Fig.4\footnote{Interestingly, 
for quasi-linear clusters, one can see that cancellation between 
the linear and second order
ISW effects can occur and  
the net temperature anisotropy can vanish near the center
if the perturbation is compensated at linear order.     
}. 
These changes lead to a temperature decrement (increment) in the 
line-of-sight to the neighbourhood of the center (edge) of the 
perturbation regardless of the sign of the fluctuation as shown in
figure 2.

For non-compensated quasi-linear perturbations, both of the 
features appear. As shown in figure 2, for overcompensated
quasi-linear voids, we would observe a hot ring that 
surrounds a prominent cold spot. 
In contrast, for undercompensated quasi-linear clusters,
we would see a cold ring that surrounds a cold spot.  

Thus it is important to study locally overcompensated (linear/quasi-linear)
voids in order to explain the origin of the Cold Spot because it
seems to have a hot ring \citep{IST10}. 

\section{Statistical Significance}
In order to estimate the statistical significance
of our model, we use CMB 
temperature anisotropy $\Delta T_f$ that is 
filtered by a spherical top-hat type
compensating filter defined as 
\BE
W_{th}(\theta;\theta_{in})=
\left\{
\begin{array}{ll}
  1 & (\theta<\theta_{in}) \\
\!  -1 & (\theta_{in}\le 
\theta \le \theta_{out}),  \\
\end{array}
\right.
\EE 
where $\theta_{out}=\cos^{-1}{(2 \cos\theta_{in}-1)}$.

In the harmonic space, the variance of filtered temperature 
$\Delta T_f$ can be written in terms of  
the angular power spectrum $C_l$ as 
\BE
\sigma^2 =A(\theta_{in})^{-2}\sum_lC_l W_l^2,
\EE 
where  $A(\theta_{in})=2 \pi (1-\cos{\theta_{in}})$ and
\BEA
W_l&=&\f{\sqrt{\pi(2l+1)}}{l+1} \biggl[2\Bigl(-x_{in}
P_l(x_{in})+P_{l-1}(x_{in})\Bigr)
\nonumber
\\
&+&x_{out}P_l(x_{out})-P_{l-1}(x_{out})
\biggr],
\nonumber
\\
x_{in}&=&\cos\theta_{in},~~ x_{out}=2x_{in}-1,
\EEA
where $P_l$ is the Legendre function. 

For detail, see appendix of \citet{IST10}.  
Using the WMAP 7-year data, a deviation of 
$\Delta T_f(\textrm{CS})=-72 \mu \textrm{K}$ is found at smoothing 
scale  $\theta_{in}=12^\circ$ if centered at 
the center of the Cold Spot \citep{IST10}. 
This corresponds to a posteriori significance level of 
$4 \sigma $ in the concordant $\Lambda$CDM model, since
$1 \sigma(\theta_{in}=12^\circ )=18\mu\textrm{K}$ for
a single realisation.

Firstly, we assume that a presence of an underdense region  
in line-of-sight to the Cold Spot contributes $18 \mu \textrm{K}$ deviation 
($1 \sigma$) of the filtered temperature at smoothing scale
$\theta_{in}=12^\circ$ as an ansatz. 
The effects of temperature anisotropy due to 
fluctuations at the last scattering surface or its neighbourhood such as 
the ordinary Sachs-Wolfe effect, the early integrated Sachs-Wolfe effect
and the Doppler effect will be considered in next section.

In order to assess the effect of non-compensation of the surrounding mass, 
we need to fix shape parameters $(r_0, r_1, r_2)$, degree of overcompensation 
$\varepsilon$, and redshift $z$ of the center. In terms of comoving
distance $d_p(z)$ to the center of the perturbation, the subtending angles of 
shape parameters are
given by $\theta_i=\arctan (r_i/d_p(z)), i=0,1,2$. 
For simplicity, we consider four types of models by fixing 
the outer boundary $r_2$ or $\theta_2$, 
the next inner boundary $r_1$ and $\varepsilon$ as shown in 
table 1. 
\begin{table*}
\centering
\begin{minipage}{150mm}
\caption{Parameters of top-hat type underdense region}
\label{tab1}
\begin{tabular}{l l l l l}
\hline
Type & $\varepsilon$ & $r_1$& $\theta_2$ & category\\
\hline
A & 0 & $(r_0+r_2)/2$ & $30^\circ$ & compensated \\
B & 0.5 & $(r_0+r_2)/2$ & $30^\circ$ &overcompensated  \\
C & 1 & $(r_0+r_2)/2$ & $30^\circ$ & overcompensated\\
D & 0 & $r_2$ & $30^\circ$ &moderately undercompensated \\
\hline
\end{tabular}
\end{minipage}
\end{table*}
Then we calculate the linear density contrast
$\delta^L_0(z)$ 
at the center that will give
a filtered temperature anisotropy $18 \mu \textrm{K}$ as a function
of the inner most boundary $r_0$ and redshift $z$.   
In order to assess the probability of a single realisation of 
such a linear perturbation, we divide the obtained amplitude of 
the linear density contrast 
by the standard deviation $\sigma_\delta^L(z)$ 
of a top-hat type matter density perturbation
expected in the concordant $\Lambda$CDM model we defined. 
We use an analytic form of the CDM transfer function \citep{Bardeen86} and 
that of the growth factor for the Friedmann-Robertson-Walker
background with dust and a cosmological constant.
 
As shown in Fig. 5, these 4 types of models give similar values for 
the best fitting parameters $\theta_0= 14^\circ-15^\circ $ and
$z\sim 1.0$. 
Provided that the perturbation is at linear regime,  
the temperature anisotropy due to the linear ISW effect 
is proportional to $\delta^L_0 r_0^3$
\citep{Inoue06}. Therefore, objects at higher redshifts will give 
larger amplitudes in the temperature anisotropy for a fixed density
contrast and a subtending angle in the accelerated epoch.  
However, if the redshift is high enough to decelerate the cosmic
expansion, then the linear ISW effect is suppressed. 
Thus the presence of peak at $z\sim 1$ can be explained as 
a result of these two different effects. As long as the perturbation is
at the linear regime, the place of the peak does not change because the
temperature anisotropy is proportional to a density contrast at the
center $\delta^L_0$. 

Although, a hot ring 
can be produced by a massive wall ($\varepsilon>0$) that 
overcompensates the underdense region inside,
overcompensated models (type B and C)
are less probable in comparison with a compensated model (type A). 
For instance, the minimum values (most probable) for type A, B, and C
are $5.6 \sigma_\delta^L, 6.1 \sigma_\delta^L,$ and $6.8 \sigma_\delta^L$. 
For overcompensated  models, as we have seen in the previous section,  
the height of the gravitational potential at the center decreases
if the sign of the potential at the wall is negative (Fig. 4) provided that the linear density contrast at the center is fixed. 
Therefore, redshift effect 
near the center becomes much 
weaker and blueshift effect at the surrounding massive 
wall suppresses further 
the overall redshifts of the CMB photons that pass through the
perturbed region. 
Thus the linear ISW effect for overcompensated
models is suppressed. In contrast, the height of the 
gravitational potentials for 
undercompensated models or
moderately undercompensated models (such as type D) 
increases in comparison with the other compensated 
models with a thin wall (such as type A) since the height of the
potential at the center increases. Therefore, the ISW effect is enhanced for
undercompensated models.
\begin{figure}
\label{figure6}
 \begin{tabular}{cc}
  \begin{minipage}{0.5\hsize}
   \begin{center}
    \includegraphics[width=4cm,clip]{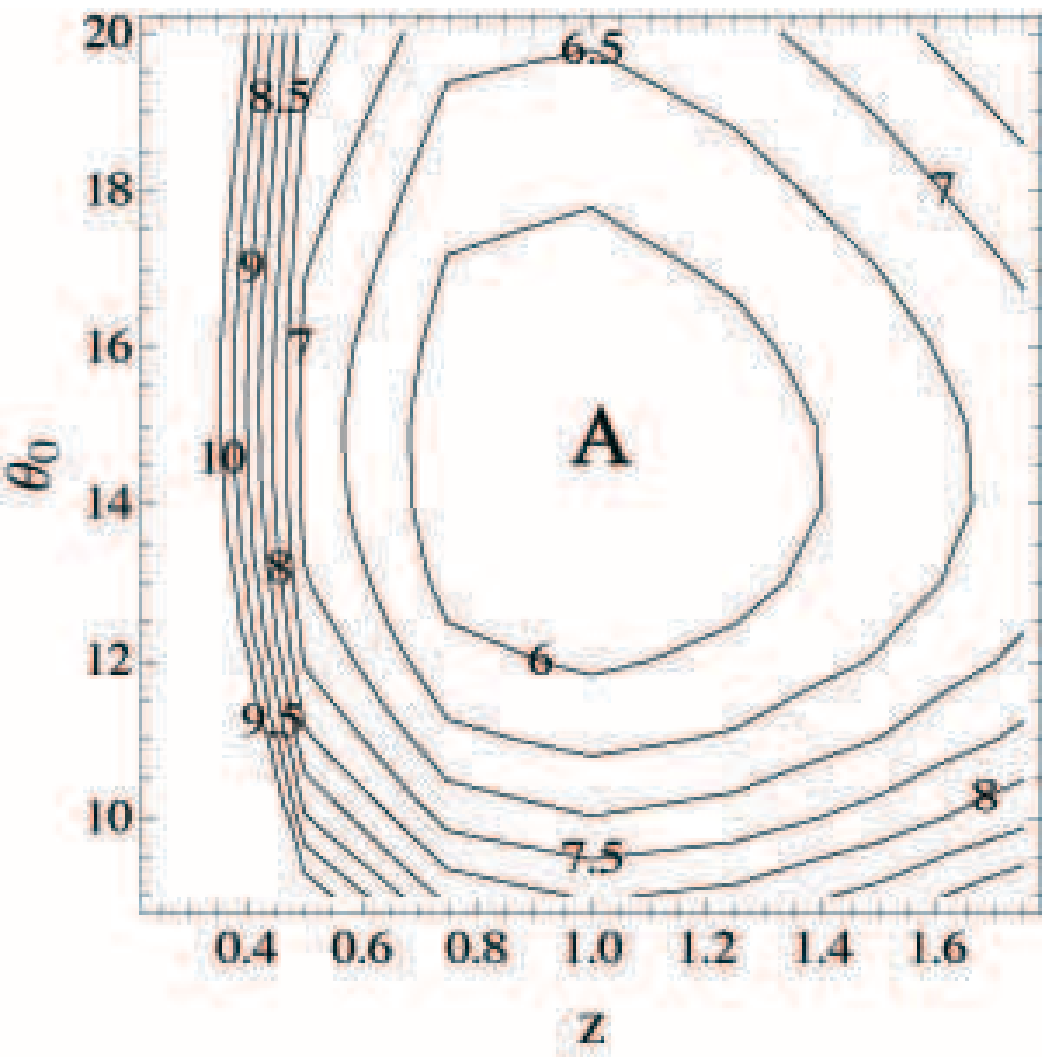}
    \includegraphics[width=4cm,clip]{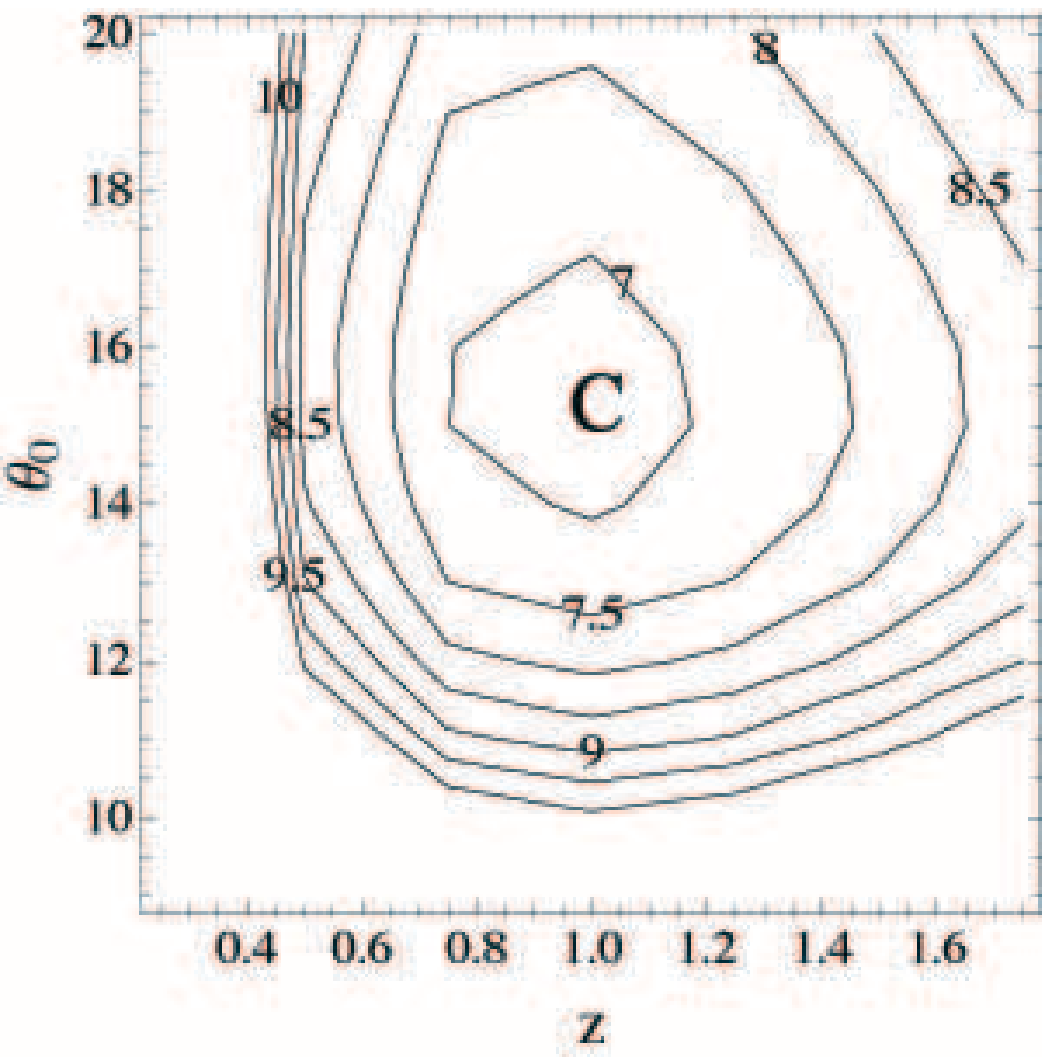}
   \end{center}
  \end{minipage}
  \begin{minipage}{0.5\hsize}
   \begin{center} 
    \includegraphics[width=4cm,clip]{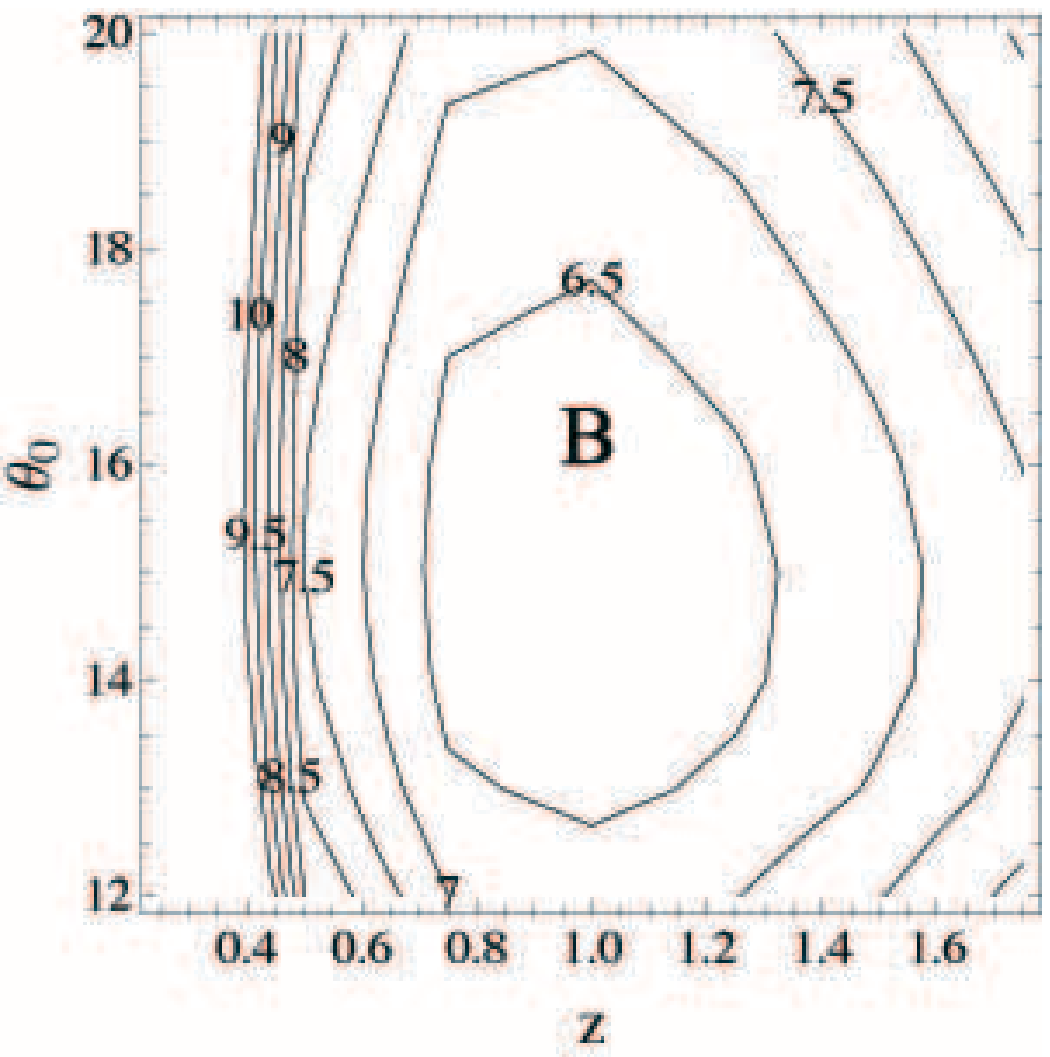}
    \includegraphics[width=4cm,clip]{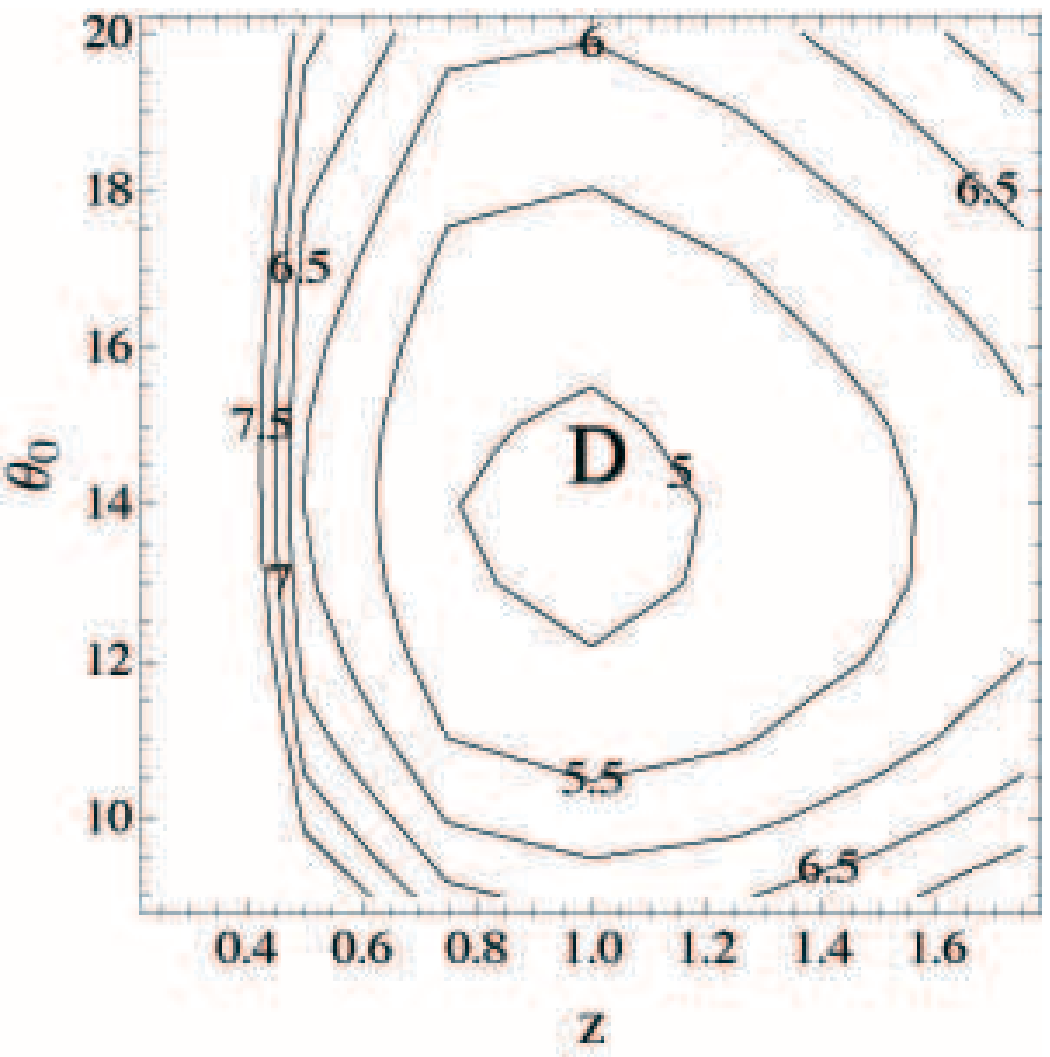}
    \end{center}
  \end{minipage}
 \end{tabular}
\vspace{0cm}
 \caption{Significance of realisation of a single top-hat type underdense perturbation that will
give a signal $\Delta T_f=-18\mu \textrm{K}(1\sigma)$ for type A(upper
 left), type B(upper right), type C(lower left), and type D(lower
 right). Contour maps of the amplitude of the linear
 density contrasts $\delta^L_0$ at the center 
in unit of the standard deviation $\sigma_\delta^L(z)$ in a concordant
 $\Lambda$CDM
model are plotted as a function of subtending angle of the inner
 boundary $\theta_0$ and redshift $z$ of the perturbation.   }
\end{figure} 

Second, in order to estimate the statistical significance of our 
single void (underdense region) model in the concordant $\Lambda$CDM
model,  we need to estimate 
the CMB anisotropy due to other effects such as the ordinary
Sachs-Wolfe effect due to spatial fluctuation of gravitational 
potentials and temperature at the last scattering. 
As we have seen in previous section, the most probable 
redshift of the underdense region is $z \sim 1$. 
Therefore, we divide contribution
from the ISW effect into two categories, `early' and `late' ones.  
In what follows, we call the ISW effect due to fluctuations of gravitational 
potential at $z>1.5$ the `early' type and at $z<1.5$ the `late' type, 
respectively. The origin of the early type ISW effect is a decay of 
potential due to residual radiation at the time of 
the last scattering. 

For simplicity, we also assume that 
the late ISW effect mainly comes from a single 
void (underdense region) in the line-of-sight to the CS.
In this calculation, 
the normalization of the angular power $C_l$ is 
set by $\sigma_8=0.80$.  In order to calculate the late type
ISW effect, we numerically solve a second-order differential equation
for the Newtonian potential $\Phi$ on large angular scales.  
We also use CMBFAST \citep{Seljak96}
for calculating the overall small-angle power 
in which the effect of acoustic
oscillation is non-negligible. However, we neglect the effect of 
reionisation since we are mainly interested in large angular power.       
Once the angular power $C_l$ of the CMB anisotropy is obtained,  
the standard deviation $\sigma=\sqrt{\langle|\Delta T_f|^2} \rangle $ 
of the filtered temperature anisotropy can be calculated straightforwardly
using equation (12). As shown in Fig. 6, the overall
amplitude (rms) of the filtered temperature is 
$\Delta T_f=18-25\mu$K for $\theta_{in}=4^\circ-16^\circ$. Because of 
anti-correlation between fluctuations due to the OSW effect and
those due to the early type ISW effect, the overall amplitude at large
angular scales is smaller than the sum of the two kinds of amplitudes.
In contrast, correlation between fluctuations due to the OSW effect
and those due to the late type ISW effect is small if the corresponding size of
the fluctuations are sufficiently smaller than the present horizon
scale. Contribution from the late type ISW effect to the overall
amplitude is about 25 per cent for $\theta_{in}=12^\circ$. 
On smaller angular scales, the contribution becomes smaller. 
For instance,  we find that 
contribution from fluctuations at redshift $0.45 < z <
0.75 $ is just $\sim 1\mu$K. This result is consistent with the previous 
calculation for super-structures in the SDSS-LRG catalog \citep{IST10}.     
\begin{figure}
\begin{center}
   \label{figSW}
   \hspace{-0.4cm}
    \includegraphics[width=8cm,clip]{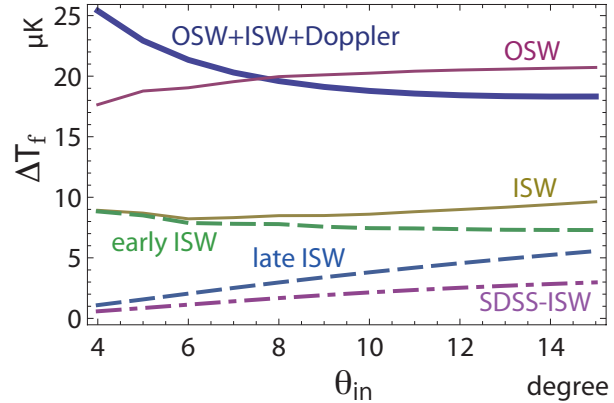}
   \label{figure7}
\end{center}
 \caption{Root-mean-square values of the amplitude of the 
filtered CMB temperature anisotropy in a
concordant $\Lambda$CDM model due to the Doppler effects (blue thick curve), 
the OSW effect (purple thin curve), the ISW effect (green thin curve),
the early type ISW effect (green dashed curve), the late type ISW effect (blue
 dashed curve), and the ISW effect from a region at $0.45<z<0.75$
 (purple dot-dashed curve).     }
\end{figure} 
\begin{figure}
\begin{center}
   \hspace{-0.4cm}
    \includegraphics[width=9cm,clip]{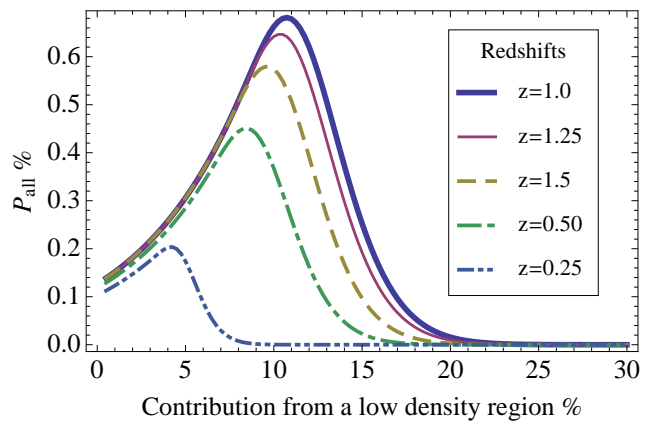}
   \label{figure8}
\end{center}
 \caption{Overall probabilities as a function of contribution from a low
 density region (void) for various redshifts. }
\end{figure} 

Finally, in order to calculate the statistical significance of a
single void (underdense region) in the line-of-sight to the CS, we 
assume that it is described by a moderately
undercompensated top-hat type linear perturbation
with $\theta_1=\theta_2=30$ degree (type D). We calculate the innermost radius
of the perturbation $r_0$ that is most probable for a given redshift 
$z$ of the center for a given amplitude of the filtered temperature $\Delta
T_f(\textrm{void})$ for $\theta_{in}=12^\circ$ as shown in the previous section.
Then for each spherical shell with 
comoving radius $d_p(z)-r_0<r<d_p(z)+r_0$, the probability
$P_{\textrm{void}}(\Delta T_f(\textrm{void});z)$
of finding
such a fluctuation is calculated based on linear perturbation 
theory assuming the CDM power spectrum (Bardeen et al. 1986).
We approximate the probability distribution of the filtered
temperature by a Gaussian distribution since the filtered temperature
is described as a sum of a number of independent modes in the harmonic space.   
The number of independent perturbed low density region for a given
angular scale $r_0/d_p(z)$ is approximated as
$N\sim d_p(z)^2/(2 r_0^2)$. Then the probability distribution function
of the filtered temperature fluctuation due to a
single void is $P_{\textrm{void}}=1-F(x)^N$
where $F(x)$ is the cumulative distribution for the Gaussian
distribution $g(x)=2/\sqrt{ 2 \pi} \exp(-x^2/2)$ and 
$x=\Delta T_f({\textrm{void}})/\sqrt{\langle (\Delta T_f({\textrm{void})})
^2\rangle }$.
The probability distribution function $P_{\textrm{LSS}}$ of finding filtered temperature
anisotropy $|\Delta T_f(\textrm{LSS})|=(72-|\Delta T_f(\textrm{void}|)) 
\mu \textrm{K}$ for $\theta_{in}=12^\circ$ due to the OSW, early type ISW
and Doppler effects can be calculated as $P_{\textrm{LSS}}=g(y)$, where 
$y=\Delta T_f({\textrm{LSS}})/\sqrt{\langle (\Delta T_f({\textrm{LSS})})^2\rangle }$.

Then the overall probability distribution function of finding such
configuration is given by $P_{\textrm{all}}=P_{\textrm{LSS}}P_{\textrm{void}}$. 
Note that we have neglected any contributions
due to fluctuations at redshifts
$z<1.5$ except for the assumed low density region (void). This is
verified if the center of the low density region is at $z\sim 1$
because the late type ISW effect is maximum at this redshift 
for the assumed angular scale $\theta_0\sim 14^\circ$ as we have seen in
the previous section.

In Fig.7, the overall probability $P_{\textrm{all}}(\ge |\Delta T_f
(\textrm{CS})|=72 \mu\textrm{K})$ 
as a function of percentage of contribution from an assumed low density region
(void), $\Delta T_f(\textrm{void})/72 \mu \textrm{K} \times 100$, is
plotted for various redshifts of the region. 
We find that the most probable values of redshift 
and contribution to the filtered temperature from a low density region
are $z =1.0$ and 11 per cent, respectively. The remaining 89
per cent of temperature anisotropy is produced by an
overdense and surrounding underdense region 
at the last scattering surface in the line-of-sight to
the CS.  The maximum value of
the overall probability of our model is $P_{\textrm{all}}=0.68$ per cent. This result
indicates that a contribution from the late type ISW effect due to an 
underdense region in the line-of-sight to the CS is not completely 
negligible. Indeed, it turns out that the possibility of 
having such a contribution is maximally 7 times larger in
comparison with the case without 
a single underdense region in the line-of-sight to the CS.

The most probable parameters of our model is
summarized in table \ref{tab2}. The temperature profile due to the most
probable underdense region is plotted in Fig.8. The expected temperature
decrement in the direction to the center is $-21\mu$K. The density
contrast is $\delta_0^L=-0.0085$ corresponding to 
$2.1 \sigma$ or a probability of 1 in 10 for a single realisation of
negative density fluctuation at $z=1$. Therefore, it is not rare to 
have such an underdense region. As we can see from Fig.8, a hot-ring
like structure around the cold region cannot be attributed to the
low density region at $z\sim 1$. If it is attributed to the OSW effect, 
we need to assume a potential hill at the last scattering surface. 
Therefore, it is most likely that 
a low density region surrounding a high density region resides at the last scattering surface.

\begin{table*}
\label{tab2}
\hspace{-2cm}
\begin{minipage}{150mm}
\caption{The most probable parameters for the type D underdense region }
\begin{tabular}{ccccccc}
\hline
$z$ & $\delta^L_{0}$ & $\theta_0$ & $\theta_1$ & $r_0$ & $\Delta T_f(\textrm{void})$ 
&$P_{\textrm{all}}(\%)
$ \\
\hline
1.0 & -0.0085& $14^\circ$ & $30^\circ$ & 570$h^{-1}$Mpc  & $7.7\mu$K & 0.68 \\
\hline
\end{tabular}
\end{minipage}
\end{table*}
\begin{figure}
\begin{center}
    \hspace{-0.5cm}
    \includegraphics[width=8cm,clip]{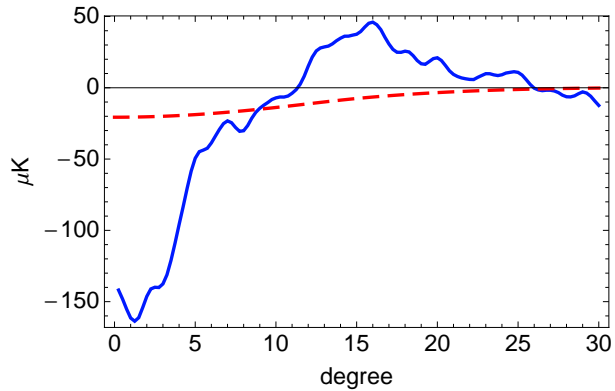}
   \label{figure9}
\end{center}
\vspace{-0.3cm}
 \caption{Temperature anisotropy due to a decaying potential of a 
single low density region (model D) (red dashed curve) 
and the observed Cold Sp
ot (WMAP7) smoothed in 
azimuthal direction (blue full curve). For detail, see \citet{IST10}. The horizontal line represents
the angular radius from the center of the CS. The parameters of
the low density region are given in table 2. }
\end{figure} 

\section{Conclusion}
We have explored a possible role of an underdense region (void)
that may cause an anomalous Cold Spot (CS) 
in the Cosmic Microwave Background (CMB) sky
located at the southern Galactic hemisphere.
Although we have found that the observed cold spot
with a hot ring can be reconstructed by an 
underdense region surrounded by a massive wall that overcompensates the
mass deficiency inside, the 
CMB temperature decrement in the line-of-sight to the 
center due to the linear integrated Sachs-Wolfe (ISW)
effect is suppressed because of blueshift of CMB 
photons at the wall surrounding an underdense region. Therefore, 
undercompensated models give better agreement with
the observed data in comparison with 
overcompensated or compensated models.
It is likely that 
$\sim$90 per cent of the CMB fluctuation is generated at an overdense 
region surrounded by an underdense region at the 
last scattering surface, and the remaining $\sim$10 per cent 
is produced due to a single underdense region that is 
moderately undercompensated with radius $r\sim 6\times
 10^2~ h^{-1}$Mpc and a density contrast 
$\delta_m\sim -0.009$ (corresponding to $\sim 2\sigma$ for a 
single realisation) at redshift $z\sim 1$. 
The probability of having such aligned structures is $\sim 0.7$ per cent if
the underdense region is moderately undercompensated. 
The temperature decrement due to the ISW effect 
is expected to be
$\sim -20 \mu \textrm{K}$ in the line-of-sight to the center.
It should be emphasised that the chance of having such an underdense
region is 7 times larger in
comparison with the case without any underdense regions 
in the line-of-sight to the CS in the concordant $\Lambda$CDM model
with Gaussian primordial perturbations. 
If such a linear underdense region does not reside 
at redshift $z \sim 1$, then either  
an abnormally large quasi-linear void at low redshift $z<0.3$, 
or non-Gaussian density fluctuations at $z \gg 1$
is necessary for explaining the origin of the CS. 
Therefore, it is very important to probe the density profile 
in the line-of-sight to the CS at redshifts $0.3<z<2$ subtending an 
angular radius $\theta>15^\circ$ in future survey. 
Ongoing projects such as LSST \citep{Ivezic08} and SKA \citep{Blake04} 
might be helpful for this purpose. 

We have not considered effects of non-sphericity of the density profile
and chance alignments between other large scale structures in the line-of-sight,
which are considered to be minor effects.  However, it will certainly improve
the accuracy of model prediction if we can take into account these
effects as well. These effects will be investigated in our future work.

\section{Acknowledgments}
I thank K.Tomita, N.Sakai, J. Silk and anonymous referee 
for useful comments. 
Some of the results presented 
here have been derived using the Healpix
package \citep{Gorski05}. 
I acknowledge the use of the Legacy Archive for Microwave Background
 Data Analysis (LAMBDA)(Lambda website).  
Support for LAMBDA is provided by the NASA Office of Space Science. 
This work is in part supported by a Grant-in-Aid for
Young Scientists (B)(20740146) of the MEXT in Japan.

\bibliographystyle{mn2e}
\bibliography{Inoue-EndNoteLibrary}
 



\end{document}